\documentclass[nofootinbib,preprint,aps,tightenlines,showpacs,superscriptaddress]{revtex4}
\usepackage[dvips]{graphicx}
\usepackage{dcolumn}
\usepackage{bm}
\usepackage{epsfig}

\newcommand{\beq}{\begin{equation}}
\newcommand{\eeq}{\end{equation}}

\begin{document}
\title{Dynamical Coupled-Channels Study of $\pi N \rightarrow \pi\pi N$
Reactions\footnote{Notice: Authored by 
Jefferson Science Associates, LLC under U.S. DOE Contract No. DE-AC05-06OR23177. 
The U.S. Government retains a non-exclusive, paid-up, irrevocable, world-wide 
license to publish or reproduce this manuscript for U.S. Government purposes. 
}} 

\vspace{0.5cm}
\author{H. Kamano}
\affiliation{ Excited Baryon Analysis Center (EBAC), Thomas Jefferson National
Accelerator Facility, Newport News, VA 23606, USA}
\author{B. Juli\'a-D\'{\i}az} 
\affiliation{ Excited Baryon Analysis Center (EBAC), Thomas Jefferson National
Accelerator Facility, Newport News, VA 23606, USA}
\affiliation{Department d'Estructura i Constituents de la Mat\`{e}ria
and Institut de Ci\`{e}ncies del Cosmos,
Universitat de Barcelona, E--08028 Barcelona, Spain}
\author{T.-S. H. Lee}
\affiliation{ Excited Baryon Analysis Center (EBAC), Thomas Jefferson National
Accelerator Facility, Newport News, VA 23606, USA}
\affiliation{Physics Division, Argonne National Laboratory, 
Argonne, IL 60439, USA}
\author{A. Matsuyama}
\affiliation{ Excited Baryon Analysis Center (EBAC), Thomas Jefferson National
Accelerator Facility, Newport News, VA 23606, USA}
\affiliation{Department of Physics, Shizuoka University, Shizuoka 422-8529, Japan}
\author{T. Sato}
\affiliation{ Excited Baryon Analysis Center (EBAC), Thomas Jefferson National
Accelerator Facility, Newport News, VA 23606, USA}
\affiliation{Department of Physics, Osaka University, Toyonaka, 
Osaka 560-0043, Japan}

\begin{abstract}
As a step toward performing a complete coupled-channels analysis of the 
world data of $\pi N, \gamma^* N \rightarrow \pi N, \eta N, \pi\pi N$
reactions, the $\pi N \rightarrow \pi\pi N$ reactions are investigated 
starting with the dynamical coupled-channels model developed in 
Phys. Rev. C76, 065201 (2007). The channels included
are $\pi N$, $\eta N$, and $\pi\pi N$  which has $\pi\Delta$, $\rho N$, 
and $\sigma N$ resonant components. The non-resonant amplitudes are 
generated from solving a set of coupled-channels equations with the  
meson-baryon potentials defined by effective Lagrangians. The resonant 
amplitudes are generated from 16 bare excited nucleon ($N^*$) states 
which are dressed by the non-resonant interactions as constrained by 
the unitarity condition. The data of total cross sections 
and $\pi N$ and $\pi\pi$  invariant mass distributions of 
$\pi^+ p \rightarrow \pi^+\pi^+ n, \pi^+\pi^0p$ and
$\pi^- p \rightarrow \pi^+\pi^- n, \pi^- \pi^0p, \pi^0\pi^0 n$ reactions
from threshold to the invariant mass $W = 2 $ GeV can be described to a 
very large extent. We show the importance of the coupled-channels effects 
and the strong interference between the contributions from the 
$\pi \Delta$, $\sigma N$, and $\rho N$ channels. The large interference 
between the resonant and non-resonant amplitudes is also demonstrated.
Possible future developments are discussed.
\end{abstract}
                                                                                
\maketitle

\section{Introduction}

It has been well recognized~\cite{lee-reviewa} that a coupled-channels
analysis of the data of meson production from $\pi N$, $\gamma N$, and
$N(e,e')$ reactions is needed to extract the parameters of the excited 
nucleon ($N^*$) states in the energy region above the $\Delta$ (1232) 
resonance. This has been pursued by using the K-matrix
models~\cite{manley84,manley03,giessen,bonn}, the Carnegie-Mellon-Berkeley 
(CMB) model~\cite{cmb} and the dynamical models~\cite{julich98,julich00,chiang01,chiang04,diaz06,
msl07,jlms07,jlmss08,djlss08,chen07}. Since two-pion production processes account for about
half of the total cross sections of $\pi N$ and $\gamma N$ reactions, the
$N^*$ parameters extracted from these analyses are reliable only when
the employed models are consistent with the  two-pion production 
data. As a step toward performing a complete analysis of the world data of 
$\pi N, \gamma^* N \rightarrow \pi N, \eta N, \pi\pi N$ reactions,
we investigate in this paper the $\pi N \rightarrow \pi\pi N$
reaction starting with the dynamical coupled-channels model developed in 
Ref.~\cite{jlms07} (JLMS) from fitting the $\pi N$ elastic scattering data.

A number of theoretical investigations of $\pi N \rightarrow \pi\pi N$
reactions have been performed using 
(1) tree-diagram calculations~\cite{oset85,jakel90,jakel92,jakel93,jensen97,kamano04,kamano06}, 
(2) chiral perturbation theory~\cite{bernard95,bernard97,fettes00,mobed05},
and (3)tree-diagram calculations including coupled-channels effects on 
nucleon resonances~\cite{julich06}. To see the scope of our investigation, 
it is useful to give a brief review of these previous works in the 
following subsections.

\subsection{Tree-diagram calculations}
The tree-diagram calculations are based on phenomenological Lagrangians 
and the Breit-Wigner form parametrization of nucleon resonances.  
All  $\pi^{\pm} p \rightarrow \pi\pi N$  channels on the proton target 
were investigated in Refs.~\cite{jensen97,kamano04,kamano06},
while only the $\pi^- p \to \pi^+\pi^- n$ channel was studied 
in Refs.~\cite{oset85,jakel90,jakel92,jakel93}.
These investigations covered the energy region up to invariant mass 
$W=1551$ MeV ($T_\pi = 600$ MeV) and only investigated the role of
the $\Delta(1232)$ and $N^\ast(1440)$ resonances. They could describe 
reasonably well the experimental data in the considered energy region. Attempt 
was also made to investigate the scalar-isoscalar two-pion decay process 
of $N^\ast(1440)\to N(\pi\pi)_{\text{S-wave}}^{I=0}$ of the Roper resonance.
Although the tree-diagram calculation is a convenient tool
to catch the qualitative features of the reaction processes,
it is of course not consistent with the unitarity requirements.
Furthermore, such an approach starts to break down in the 
$W \gtrsim$ 1.5 GeV region where more $N^*$ states needed to be considered.
\subsection{Chiral perturbation theory}

Chiral perturbation theory calculations of $\pi N \rightarrow \pi\pi N$
reactions have been performed up to the order $O(q^3)$.
The focus of Refs.~\cite{bernard95} was on evaluating the $\pi N\to\pi\pi N$ 
threshold amplitudes $D_{1,2}$ and extracting $\pi\pi$ scattering length 
within the heavy baryon chiral perturbation theory. This work was extended 
in Refs.~\cite{bernard97,fettes00} to also compare with the cross section 
data up to $W = 1.38$ GeV ($T_\pi = 400$ MeV). They found that the loop 
corrections are small and it is difficult to extract the isoscalar 
S-wave $\pi\pi$ scattering length because of the large uncertainty associated 
with the $N^\ast(1440)\to N(\pi\pi)_{\text{S-wave}}^{I=0}$ decay.
Similar chiral perturbation theory calculation was also performed to
compare with more data near the threshold by Mobed et al~\cite{mobed05}.

Of course, the application of chiral perturbation 
theory to investigate the $\pi N \rightarrow \pi\pi N$ reactions near
threshold is an important advance. But it is not clear how it can be 
used to investigate the nucleon resonances up to $W = 2$ GeV.

\subsection{ Tree-diagram calculation including  coupled-channels 
effects on $N^*$ }
The $\pi N\to \pi\pi N$ study performed by the Julich group~\cite{julich06}
also used the tree-diagrams generated from phenomenological Lagrangians, but 
including the $P_{33}(1232)$, $P_{11}(1440)$, $D_{13}(1520)$, $S_{11}(1535)$, 
and $S_{11}(1650)$ resonance states. They focused on the low $ W \le $ 1.38
GeV ($T_\pi = 400$ MeV) region. The coupled-channels effects 
are included by using the non-resonant amplitudes generated from
the $\pi N$ coupled-channels model of Ref.~\cite{julich00} to evaluate
the self-energy  and the decay functions of 
the considered resonances.  They could describe to a large
extent most of the available data in the $ W < 1.38 $ GeV region.
They found that their calculations started to break down in the 
higher $W$ region.

\subsection*{}
In this work, we depart from all of the earlier works described above by
considering the whole energy region from threshold up to $W = 2$ GeV and
all nucleon resonances listed by Particle Data Group~\cite{pdg} (PDG). 
The calculations are performed by using the JLMS model. Schematically, 
the following coupled integral equations in each partial 
wave are solved within the JLMS model
\begin{eqnarray}
T_{\alpha,\beta}(p_\alpha,p_\beta;E)= V_{\alpha,\beta}(p_\alpha,p_\beta)
+ \sum_{\delta}
 \int p^{2}d p  V_{\alpha,\delta}(p_\alpha, p )
G_{\delta}( p ,E)
T_{\delta,\beta}( p  ,p_\beta,E)  \,,
\label{eq:teq}
\end{eqnarray}
with
\begin{eqnarray}
V_{\alpha,\beta}(p_\alpha,p_\beta)= v_{\alpha,\beta}(p_\alpha,p_\beta)+
\sum_{N^*}\frac{\Gamma^{\dagger}_{N^*,\alpha}(p_\alpha)
 \Gamma_{N^*,\beta}(p_\beta)} {E-M^0_{N^*}} \,,
\label{eq:veq}
\end{eqnarray}
where $\alpha,\beta,\delta = \gamma N, \pi N, \eta N,$ and
$\pi\pi N$ which has $\pi \Delta, \rho N, \sigma N$ resonant components,
$G_\delta (p,E)$ is the propagator of channel $\delta$, $M^0_{N^*}$
is the mass of a bare excited nucleon state $N^*$, $v_{\alpha,\beta}$
is defined by meson-exchange mechanisms, and the 
 vertex interaction $\Gamma_{N^*,\beta}$ defines the
$N^*\rightarrow \beta$ decay. The fits to $\pi N$ elastic scattering data were 
achieved by including one or two bare $N^*$ states in all 
$S$, $P$, $D$ and $F$ 
partial waves. The details can be found in Ref.~\cite{jlms07}.

While the non-resonant interactions $v_{\alpha,\beta}$ are deduced from
phenomenological Lagrangians, the model contains many parameters mainly 
due to the lack of sound theoretical guidance in parametrizing the bare
$N^* \rightarrow \pi N, \eta N, \pi\Delta, \rho N, \sigma N$ form factors.
Although fitting 
the  $\pi N$ elastic scattering data is rather complex and time consuming, 
as reported in Ref.~\cite{jlms07}, it is clearly not sufficient to determine
these parameters; in particular the parameters associated with
the unstable $\pi\Delta$, $\rho N$, and $\sigma N$ channels. Thus it is 
important to test the JLMS model in the study of 
$\pi N \rightarrow \pi\pi N$ reactions which are known to be dominated
by these channels. This is the main purpose of this work. 
The results 
presented in this paper are therefore obtained using the parameters 
taken from the JLMS model.

It is tempting to try to use the available 
$\pi N \rightarrow \pi\pi N$ data to
improve the JLMS model. However this is not an 
easy task at the present time mainly because of the lack of sufficient 
experimental data in the considered energy region up to 
$W = 2 $ GeV ($T_\pi \sim 1.6$) GeV.
Furthermore most of the data at high $W$, obtained before 1970's, are 
not accessible~\cite{arndt}. In an effort by R. Arndt~\cite{arndt}, some
of the final $\pi\pi$ and $\pi N$ invariant mass distribution data have 
been recovered and considered in this paper along with the most studied 
total cross section data~\cite{data}. This allows us to investigate these two 
observables for all possible final $\pi\pi N$ states of $\pi^\pm p$ reactions
and in the entire energy region from the threshold to $W = $ 
2 GeV. Very limited data on the angular distributions 
can be found in the literature, only few in the low $W$ region and 
practically nothing in the higher W region where we need pin down
the parameters associated with many $N^*$ states. We thus will not consider 
these observables. For the same reason it is difficult to use the available
$\pi N \rightarrow \pi\pi N$ data at present time.
This is mainly due to the problem that the minimization in determining a 
large amount of parameters of our model is heavily weighted by the very 
extensive and far more precise data of $\pi N$ elastic scattering. We thus 
focus in this paper on investigating the dynamical content of the JLMS 
model, in particular on several aspects which were not addressed before 
such as the effects of coupled-channels and the role played by the 
interference between the different channels.
Even with this limitation our investigation is clearly more extensive 
than all previous works reviewed above.

In section II, we briefly recall the formulas of Ref.~\cite{msl07}
for calculating the $\pi N \rightarrow \pi\pi N$ amplitudes within the JLMS
model. 
To give more details about our calculations, explicit expressions 
for calculating the total cross sections and $\pi N$ and $\pi\pi$ 
invariant mass distributions are given in section III.
The results are presented  in section IV. In section V, we give a
summary and discuss 
possible future developments.
\begin{figure}[t]
\centering
\includegraphics[clip,width=12cm]{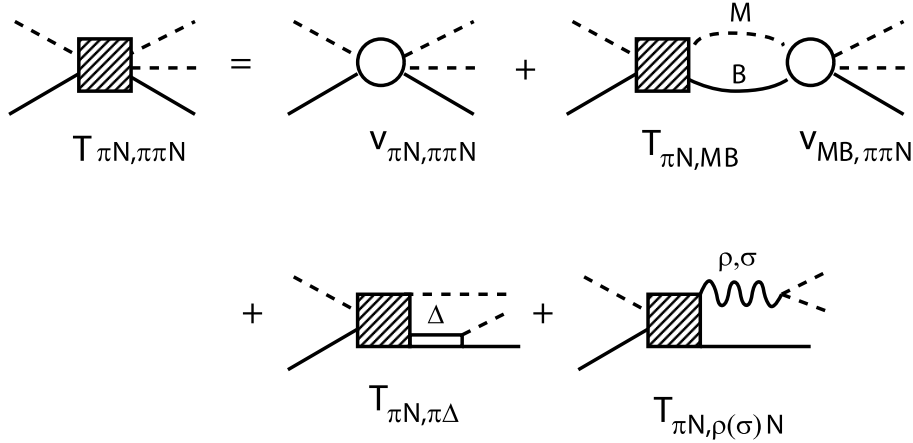}
\caption{Graphical representations of  $T_{\pi N, \pi\pi N}$ of 
Eqs.(\ref{eq:tpipin-1})-(\ref{eq:tpipin-sigman}).}
\label{fig:mb-pipin}
\end{figure}

\section{Formulation}
 
Within a Hamiltonian formulation~\cite{msl07} within which the JLMS model
was developed, the
$\pi N \rightarrow \pi\pi N$ amplitude 
is illustrated in Fig.\ref{fig:mb-pipin} and can be written as
\begin{eqnarray}
T_{\pi N,\pi \pi N}(E) = 
 T^{\text{dir}}_{\pi N,\pi\pi N}(E)
+T^{\pi\Delta}_{\pi N,\pi \pi N}(E)
+T^{\rho N}_{\pi N,\pi\pi N}(E)
+T^{\sigma N}_{\pi N, \pi \pi N}(E),
\label{eq:tpipin-1}
\end{eqnarray}
with
\begin{eqnarray}
T^{\text{dir}}_{\pi N,\pi \pi N}(E)
&=&
 v_{\pi N,\pi \pi N}+
\sum_{MB} T_{\pi N,MB}(E)G_{MB}(E) v_{MB,\pi\pi N}
\label{eq:tpipin-dir} \,, \\
T^{\pi \Delta}_{\pi N,\pi\pi N}(E)
&=&
T_{\pi N, \pi \Delta}(E)
G_{\pi\Delta}(E)
\Gamma_{\Delta \rightarrow \pi N}
\label{eq:tpipin-pid} \,, \\
& & \nonumber \\
T^{\rho N}_{\pi N, \pi\pi N}(E)
&=& 
T_{\pi N,\rho N}(E) G_{\rho N}(E) 
h_{\rho\rightarrow \pi\pi} 
\label{eq:tpipin-rhon} \,, \\
& & \nonumber \\
T^{\sigma N}_{\pi N,\pi \pi N}(E)
&=&
T_{\pi N,\sigma N}(E) G_{\sigma N}(E)
h_{\sigma\rightarrow \pi\pi},
\label{eq:tpipin-sigman}
\end{eqnarray}
where $\Gamma_{\Delta\rightarrow \pi N}$, $h_{\rho\rightarrow \pi\pi}$, and
$h_{\sigma\rightarrow \pi\pi}$  describe the  
$\Delta(1232)\rightarrow \pi N$, $\rho\rightarrow \pi\pi$, and 
$\sigma \rightarrow \pi\pi$ decays, respectively, $G_{MB}(E)$ for 
$MB=\pi N,\eta N,\pi\Delta, \rho N, \sigma N$ are the meson-baryon propagators.
The $\pi N \rightarrow MB$ transition amplitudes 
are
\begin{eqnarray}
T_{\pi N,MB}(E)  &=&  t_{\pi N,MB}(E)
+ t^R_{\pi N,MB}(E) \,.
\label{eq:tmbmb}
\end{eqnarray}
The second term in the right-hand-side of
Eq.(\ref{eq:tmbmb}) is the resonant term defined by
\begin{eqnarray} 
t^R_{MB,M^\prime B^\prime}(E)= \sum_{N^*_i, N^*_j}
\bar{\Gamma}_{MB \rightarrow N^*_i}(E) [D(E)]_{i,j}
\bar{\Gamma}_{N^*_j \rightarrow M^\prime B^\prime}(E) \,,
\label{eq:tmbmb-r} 
\end{eqnarray}
with
\begin{eqnarray}
[D(E)^{-1}]_{i,j}(E) = (E - M^0_{N^*_i})\delta_{i,j} - \bar{\Sigma}_{i,j}(E)\,,
\label{eq:nstar-g}
\end{eqnarray}
where $M_{N^*}^0$ is the bare mass of the excited nucleon state $N^*$, and
the self-energies are 
\begin{eqnarray}
\bar{\Sigma}_{i,j}(E)= \sum_{MB}\Gamma_{N^*_i\rightarrow MB} G_{MB}(E)
\bar{\Gamma}_{MB \rightarrow N^*_j}(E) \,.
\label{eq:nstar-sigma}
\end{eqnarray}
The dressed vertex interactions in Eq.(\ref{eq:tmbmb-r}) and
Eq.(\ref{eq:nstar-sigma}) are 
 (defining
 $\Gamma_{MB\rightarrow N^*}=\Gamma^\dagger_{N^* \rightarrow MB}$)
\begin{eqnarray}
\bar{\Gamma}_{MB \rightarrow N^*}(E)  &=&  
  { \Gamma_{MB \rightarrow N^*}} + \sum_{M^\prime B^\prime}
t_{MB,M^\prime B^\prime}(E)
G_{M^\prime B^\prime}(E)
\Gamma_{M^\prime B^\prime \rightarrow N^*}\,, 
\label{eq:mb-nstar}
\\
\bar{\Gamma}_{N^* \rightarrow MB}(E)
 &=&  \Gamma_{N^* \rightarrow MB} +
\sum_{M^\prime B^\prime} \Gamma_{N^*\rightarrow M^\prime B^\prime}
G_{M^\prime B^\prime }(E)t_{M^\prime B^\prime,M B}(E) \,. 
\label{eq:nstar-mb}
\end{eqnarray}
The non-resonant amplitudes $ t_{MB,M^\prime B^\prime}$ 
in Eq.(\ref{eq:tmbmb}) and Eqs.(\ref{eq:mb-nstar})-(\ref{eq:nstar-mb})
are defined by the following coupled-channels equations
\begin{eqnarray}
t_{MB,M^\prime,B^\prime}(E)&= &v_{MB,M^\prime B^\prime}(E) 
+ \sum_{M^{\prime\prime}B^{\prime\prime}}
v_{MB,M^{\prime\prime}B^{\prime\prime}}(E)
G_{M^{\prime\prime}B^{\prime\prime}}(E) 
t_{M^{\prime\prime}B^{\prime\prime},M^\prime B^\prime}(E).
\label{eq:cc-mbmb}
\end{eqnarray}

\begin{figure}[t]
\centering
\includegraphics[clip,width=16cm]{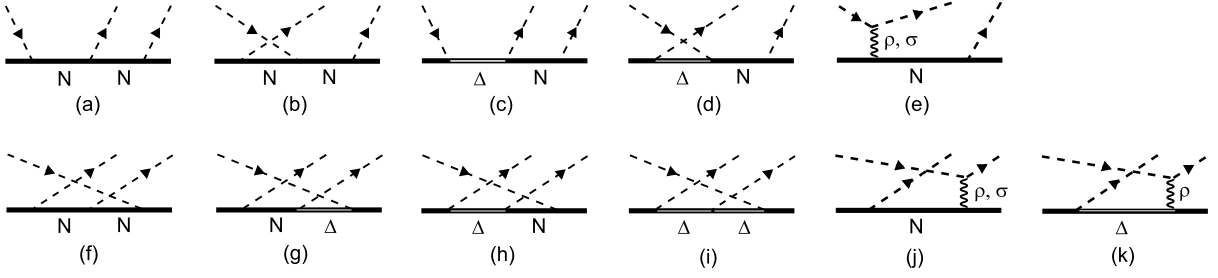}
\caption{The considered $v_{\pi N,\pi\pi N}$. }
\label{fig:mbpipin}
\end{figure}
The channels included are $MB =\pi N,\eta N, \pi\Delta, \rho N, \sigma N$.
All quantities defined above are described in detail in
Refs.~\cite{msl07,jlms07} 
and can be calculated within the JLMS model. The only exception is
the direct production term $v_{\pi N,\pi\pi N}$ in Eq.(\ref{eq:tpipin-dir}) 
which is not included in the JLMS model. The procedure for deriving 
$v_{\pi N,\pi\pi N}$ from Lagrangians by using the method of unitary 
transformation is explained in Ref.~\cite{msl07}. In this work we 
consider a $v_{\pi N,\pi\pi N}$ model which involves only the
$N$ and $\Delta$(1232) intermediate baryon states
such that its parameters can also be consistently taken from the JLMS 
model. Our calculations thus do not have any adjustable parameters.
The  mechanisms of the considered $v_{\pi N,\pi\pi N}$
are illustrated in Fig.\ref{fig:mbpipin}. 

The calculations of the terms $T^{MB}_{\pi N,\pi\pi N}$ with 
$MB=\pi\Delta, \rho N, \sigma N$, defined 
by Eqs.(\ref{eq:tpipin-pid})-(\ref{eq:tpipin-sigman}), are 
straightforward. On the other hand, the calculation of the second 
term of $T^{\text{dir}}_{\pi N,\pi\pi N}$, 
defined by Eq.(\ref{eq:tpipin-dir}), 
is much more complex. To simplify the calculation, we first note that 
the mechanisms (a)-(e) in the upper row of Fig.~\ref{fig:mbpipin} can be written as
\begin{eqnarray}
v^{(a-e)}_{\pi N, \pi\pi N}\sim v_{\pi N, \pi N} G_{\pi N}(E) h_{N\rightarrow\pi N},
\label{eq:dir-app-1}
\end{eqnarray}
where $v^{(a-e)}_{\pi N, \pi\pi N}$ is the sum  of all 
mechanisms $(a)$ - $(e)$, $v_{\pi N, \pi N}$ is part of the driving term 
in calculating the amplitude $T_{\pi N,\pi N}(E)$ from JLMS model, 
and $h_{N\rightarrow\pi N}$ is the $N\rightarrow \pi N$ vertex function. 
If we neglect the final state interactions on the mechanisms 
(f)-(k) of Fig.\ref{fig:mbpipin}, we can write
\begin{eqnarray}
T^{\text{dir}}_{\pi N,\pi\pi N} &\sim& v^{(f-k)}_{\pi N,\pi\pi N} +
[v_{\pi N,\pi N} + \sum_{MB}T_{\pi N,MB} G_{MB}(E)v_{MB,\pi N}(E) ] 
G_{\pi N}(E) h_{N\rightarrow\pi N} \nonumber \\
&\sim& v^{(f-k)}_{\pi N,\pi\pi N} + T_{\pi N,\pi N}(E)G_{\pi N}(E) 
h_{N\rightarrow\pi N}.
\label{eq:dir-app-2}
\end{eqnarray}
In deriving the above equation, we have used the full scattering 
equation $T = V + TGV$ of the JLMS model with the approximation that 
only  the $\Delta$ (1232) bare state and the $\pi N$ state are kept 
in summing the intermediate states.  We use Eq.(\ref{eq:dir-app-2}) in this work.
 \section{Cross section formula}
For presenting results, we here give explicit formula for calculating
the cross sections of $\pi N \rightarrow \pi\pi N$ reactions. 
Within the formulation
of Refs.~\cite{msl07,jlms07},
the S-matrix is defined by
\begin{eqnarray}
S_{fi} = \delta_{fi} - 2\pi {\it i} \delta^4 (P_f-P_i) \langle f|T|i\rangle  \,,
\label{eq:s-mtx}
\end{eqnarray}
and the plane-wave state is normalized as
\begin{eqnarray}
\langle \vec{q}|\vec{p}\rangle  =\delta (\vec{q}-\vec{p} )\,.
\end{eqnarray}
In the center of mass frame, the momentum variables of
the $\pi N \rightarrow \pi\pi N $ reaction with invariant mass  $W$
can be specified as
\begin{eqnarray}
a(\vec{p}_a)+ b(\vec{p}_b)\rightarrow 
c(\vec{p}_c) + d(\vec{p}_d) + e(\vec{p}_e)
\label{eq:process}
\end{eqnarray}
where
$\vec{p}_a =-\vec{p}_b=\vec{k}$ with $k$ defined by $W=E_a(k)+E_b(k)$,
 $\vec{p}_c + \vec{p}_d =-\vec{p}_e=\vec{k'}$, and
 $(c+d+e)$ can be any possible charged states formed from two pions and
one nucleon. The total cross section of the process Eq.(\ref{eq:process}) can
then be  written as
\begin{eqnarray}
\sigma_{ab \rightarrow cde} & = & \frac{1}{v}(2\pi)^4 \int d\vec{p}_c d\vec{p}_d d\vec{p}_e
 \delta^4(p_a + p_b - p_c - p_d - p_e)
\nonumber\\
&&
\times
\frac{1}{(2s_a +1)(2s_b+1)}
\sum_{\bar{i},\bar{f}}|
\langle \vec{p}_c\vec{p}_d\vec{p}_e,f|T|\vec{p}_a\vec{p}_b, i\rangle |^2 \nonumber\\
 & = &
 \frac{E_a(k)E_b(k)}{W k}(2\pi)^4
 \int \frac{d\vec{p}_c}{E_c}\frac{d\vec{p}_d}{E_d}\frac{d\vec{p}_e}{E_e}
 \delta^4(p_a + p_b - p_c - p_d - p_e) \nonumber \\
&&\times
\frac{1}{(2s_a +1)(2s_b+1)}
\sum_{\bar{i},\bar{f}}|\sqrt{E_cE_dE_e}
\langle \vec{p}_c\vec{p}_d\vec{p}_e,f|T|\vec{p}_a\vec{p}_b, i\rangle |^2 
\label{eq:sig-tot}
\end{eqnarray}
where $i,f$ denote all spin $(s_a, s_{az})$ and isospin $(t_a, t_{az})$
quantum numbers, and $\sum_{\bar{i},\bar{f}}$ means summing over only
spin quantum numbers.
The above equation can be written as
\begin{eqnarray}
\sigma_{ab \rightarrow cde}& = & \int_{m_c+m_d}^{W-m_e}
\frac{d\sigma}{dM_{cd}}dM_{cd} 
\label{eq:eq:sigma-t}
\end{eqnarray}
with
\begin{eqnarray}
\frac{d\sigma}{dM_{cd}}
=
\frac{\rho_i}{k^2}16\pi^3
\int d\Omega_{k_{cd}} d\Omega_{k'}
\frac{k_{cd}k'}{W}
\frac{1}{(2s_a +1)(2s_b+1)}
\sum_{\bar{i},\bar{f}}|\sqrt{E_cE_dE_e}
\langle \vec{p}_c\vec{p}_d\vec{p}_e,f|T|\vec{k},i\rangle |^2 .
\nonumber\\
\end{eqnarray}
Here we have defined 
\begin{eqnarray}
\rho_i &=&\pi\frac{kE_a(k)E_b(k)}{W} \,,\nonumber \\
W      &=& E_e(k') + E_{cd}(k') \,, \nonumber \\
E_{cd}(k')&=& \sqrt{M_{cd}^2 + (k')^2}\,, \\
M_{cd} &=& E_c(k_{cd}) + E_d(k_{cd}) \,,\nonumber
\end{eqnarray}
and $\vec{k}_{cd}$ is the relative momentum between $c$ and $d$ 
in the $e$ rest frame.

The T-matrix elements in the above equation are calculated from
Eqs.(\ref{eq:tpipin-dir})-(\ref{eq:tpipin-sigman}). For
the quasi two-body processes Eqs.(\ref{eq:tpipin-pid})-(\ref{eq:tpipin-sigman})
 with a resonant unstable particle  $R$ which decays into $ c + d$ state, 
the T-matrix elements take the following form
\begin{eqnarray}
\langle {p}_c\vec{p}_d\vec{p}_e,f|T|\vec{k},i\rangle  & = & \sum_{s_{R_z},t_{R_z}}
\frac{
\langle \vec{p}_c,s_{cz},t_{cz};\vec{p}_d,s_{dz},t_{dz}|H_I|
\vec{k}',s_{Rz},t_{R_z} \rangle 
}{E - E_e(k') - E_R(k') - \Sigma_{eR}(k',W)} \nonumber \\
& &\times 
\langle \vec{k}',s_{Rz},t_{Rz};-\vec{k}',s_{ez},t_{ez}|
T|\vec{k},s_{az},t_{az};-\vec{k},s_{bz},t_{bz}\rangle .
\end{eqnarray}
For any spins and isospins and c.m. momenta $\vec{p}$ and $\vec{p}'$, 
the $MB\rightarrow M'B'$ T-matrix elements are in general defined
by
\begin{eqnarray}
&&
\langle \vec{p}',s_{M'z},t_{M'z}; -\vec{p}',s_{B'z},t_{B'z}|
T|\vec{p},s_{Mz},t_{Mz};-\vec{p},s_{Bz},t_{Bz}\rangle  \nonumber 
\\
&&= \sum_{S,S',L,L',J,M,I} Y_{L',M_L'}(\hat{p}')Y_{L,M_L}^*(\hat{p})
\nonumber\\
&& \times
\langle s_{M'},s_{M'z},s_{B'},s_{B'z}|S',S_z'\rangle \langle L',M_L',S',S_z'|J,M\rangle  
\langle t_{M'},t_{M'z},t_{B'},t_{B'z}|I,I_z\rangle \nonumber \\
& & \times
\langle s_M,s_{Mz},s_B,s_{Bz}|S,S_z\rangle \langle L,M_L,S,S_z|J,M\rangle 
\langle t_M,t_{Mz},t_B,t_{Bz}|I,I_z\rangle 
 \nonumber \\
&& \times
\langle p',S'L'J|T^I(W)|p,SLJ\rangle ,
\end{eqnarray}
where $\langle j_1,j_1,m_1,m_2|jm\rangle $ is the Clebsch-Gordan coefficient of
the $\vec{j}_1+\vec{j}_2 = \vec{j}$ coupling,
 $\langle k',S'L'J|T^I(W)|k, SLJ\rangle $ is the partial wave amplitude defined by
the  total angular
momentum $J$, orbital angular momentum $L$, total spin $S$, and total
isospin $I$. We use the JLMS model to generate
$\langle k',S'L'J|T^I(W)|k, SLJ\rangle $.

As for the $R \rightarrow cd $ vertex function, the combination
\beq
\sqrt{E_c(p_c)E_d(p_d)E_R(k')}
\langle\vec{p}_c,s_{cz},t_{cz};\vec{p}_d,s_{dz},t_{dz}|H_I|
\vec{k}',s_{Rz},t_{Rz}\rangle \nonumber
\eeq
is Lorentz invariant and therefore can be written in terms of its matrix 
element in the rest frame of
$R$
\begin{eqnarray}
&&
\langle \vec{p}_c,s_{cz},t_{cz};\vec{p}_d,s_{dz},t_{dz}|H_I|\vec{k}',s_{Rz},t_{Rz}\rangle 
\nonumber\\
&=& \delta(\vec{p}_c+\vec{p}_d - \vec{k}')
\sqrt{\frac{E_c(k_{cd})E_d(k_{cd})M_R}
{{E_c(p_c)E_d(p_d)E_R(k')}}}
\langle \vec{k}_{cd},s_{cz},t_{cz};-\vec{k}_{cd},s_{dz},t_{dz}|
H_I|\vec{0},s_{Rz},t_{Rz}\rangle ,
\end{eqnarray}
with
\begin{eqnarray}
&&
\langle \vec{k}_{cd},s_{cz},t_{cz};-\vec{k}_{cd},s_{dz},t_{dz}|
H_I|\vec{0},s_{Rz},t_{Rz}\rangle 
\nonumber\\
&&=
\sum_{L_{cd},m_{cd},S_{cd},S_{cdz}}
[\langle s_c,s_{cz},s_d,s_{dz}|S_{cd},S_{cdz}\rangle 
\langle L_{cd},m_{cd},S_{cd},S_{cdz}|s_R,s_{Rz}\rangle  \nonumber \\
& & \,\,\,\,\,\, \times\langle t_{c},t_{cz},t_{d},t_{dz}|t_R,t_{Rz}\rangle 
Y_{L_{cd},m_{cd}}(\hat{k}_{cd})
F_{L^R_{cd},S^R_{cd}}(k_{cd})].
\end{eqnarray}
The vertex functions are
\begin{eqnarray}
F_{L^\Delta_{\pi N}, S^\Delta_{\pi N}}(q) &=& if_{\Delta\rightarrow \pi N}(q)
\label{eq:ff-pid} \\
F_{L^\sigma_{\pi \pi}, S^\sigma_{\pi \pi}}(q) 
&=& \sqrt{2}f_{\sigma\rightarrow \pi\pi}(q) \label{eq:ff-sn}
\\
F_{L^\rho_{\pi \pi}, S^\rho_{\pi \pi}}(q) &=& 
(-1)\sqrt{2}f_{\rho\rightarrow \pi\pi}(q) \label{eq:ff-rn}
\end{eqnarray}
where $L^\Delta_{\pi N}=1, S^\Delta_{\pi N}=3/2$, 
$L^\sigma_{\pi \pi}=0, S^\sigma_{\pi \pi}=0$, 
$ L^\rho_{\pi \pi}=1, S^\rho_{\pi \pi}=1$.
Here it is noted that the factor $\sqrt{2}$ in 
Eqs.(\ref{eq:ff-sn})-(\ref{eq:ff-rn}) comes from the Bose symmetry
of pions, and the phase factor $i$ and $(-1)$ are chosen to be
consistent with the non-resonant interactions involving
$\pi N\Delta$, $\sigma \pi\pi$ and $\rho\pi\pi$ vertex
interactions defined in Ref.~\cite{msl07}.
The form factors in Eqs.(\ref{eq:ff-sn})-(\ref{eq:ff-rn})
\begin{eqnarray}
f_{\Delta\rightarrow \pi N}(q) &=&
- \frac{f_{\pi N\Delta}}{m_{\pi}}
\frac{1}{(2\pi)^{3/2}}\sqrt{\frac{4\pi}{3}}q
\frac{1}{\sqrt{2E_\pi(q)}}\sqrt{\frac{E_N(q)+m_N}{2E_N(q)}}
\left(\frac{\Lambda_{\pi N\Delta}^2}{\Lambda_{\pi N\Delta}^2 + q^2}\right)^2,
\\
f_{\sigma \rightarrow \pi \pi}(q) &=&
\frac{g_{\sigma\pi\pi}}{\sqrt{m_\pi}} \frac{\Lambda_{\sigma\pi\pi}^2}{\Lambda_{\sigma\pi\pi}^2+q^2},
\\
f_{\rho \rightarrow \pi \pi}(q) &=&
\frac{g_{\rho\pi\pi}}{\sqrt{m_\pi}}\frac{q}{\Lambda_{\rho\pi\pi}}
\left(\frac{\Lambda_{\rho\pi\pi}^2}{\Lambda_{\rho\pi\pi}^2+q^2}\right)^2,
\end{eqnarray}
with
$f_{\pi N\Delta}=2.049$,
$g_{\sigma\pi\pi}=0.7750$,
$g_{\rho\pi\pi}=0.6684$,
$\Lambda_{\pi N\Delta} = 649$ MeV,
$\Lambda_{\sigma\pi\pi} = 378$ MeV, and
$\Lambda_{\rho\pi\pi} = 461$ MeV.
The above vertex functions are determined from fitting the
$\pi N$ phase shifts in $P_{33}$~\cite{sl96}  and $\pi\pi$ phase 
shifts~\cite{jl86}.

With the vertex function $f_{R\rightarrow cd}(q)$ given above, 
the self-energy appearing in $eR$ Green function
 $\Sigma_{eR}$ 
($eR = \pi\Delta,N \rho , N \sigma $),
are calculated from (see Appendix \ref{app1})
\begin{eqnarray}
\Sigma_{eR}(k,E) &=& \frac{m_R}{E_R(k)}
\int q^2 dq \frac{M_{eR}(q)}{[M^2_{eR}(q)+k^2]^{1/2}}
\frac{|f_{R\rightarrow cd}(q)|^2}
{E-E_e(k)-\{[E_c(q) + E_d(q)]^2 + k^2\}^{1/2} +i\varepsilon}.
\nonumber\\
&&
\end{eqnarray}
To derive the above equation, we have used the Lorentz transformation 
to calculate the self-energy in arbitrary frame from the vertex function
defined in the rest frame of $R$.

\begin{figure}[t]
\centering
\includegraphics[clip,width=15cm]{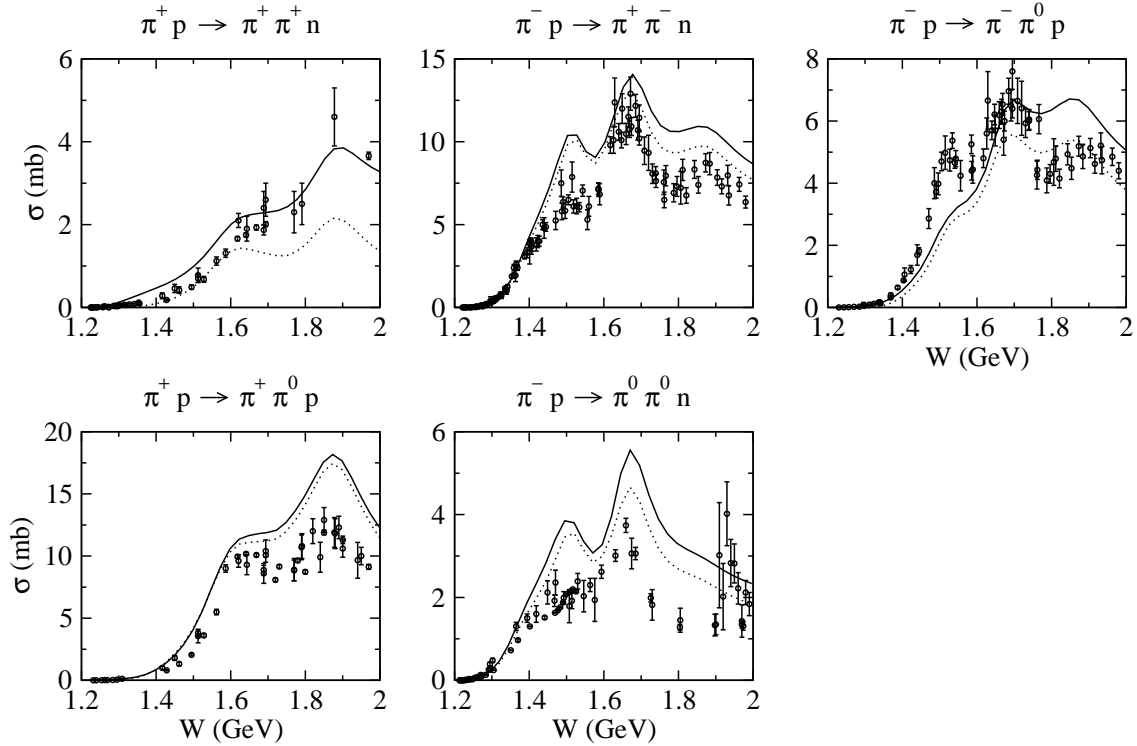}
\caption{ The total cross sections predicted
(solid curves) from the JLMS model are
compared with the data~\cite{data}. The dotted curves are from turning off
the amplitude $T^{\text{dir}}_{\pi N, \pi\pi N}$. 
}
\label{fig:jlms-pipp}
\end{figure}

\begin{figure}[t]
\centering
\includegraphics[clip,width=15cm]{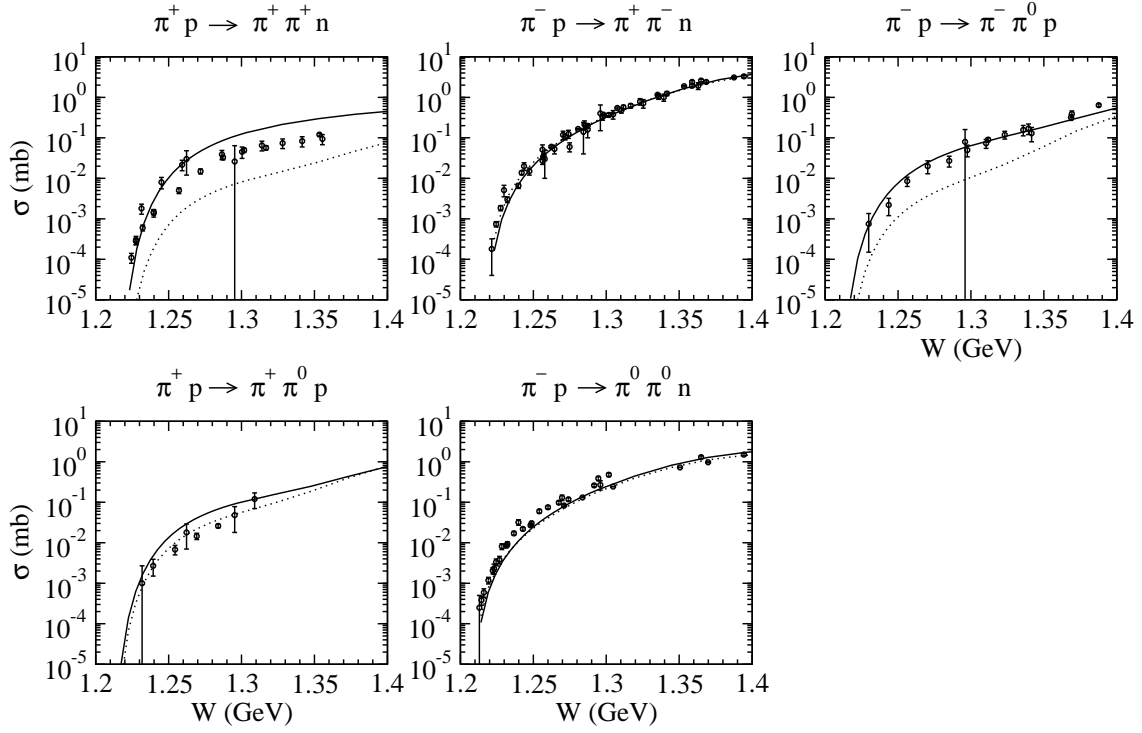}
\caption{The total cross sections 
predicted (solid curves) from the JLMS model are
compared with the data~\cite{data}. The dotted curves are from turning off
the amplitude $T^{\text{dir}}_{\pi N, \pi\pi N}$.}
\label{fig:jlms-pipp-th}
\end{figure}

\section{Results}

With the formula presented in sections II and III, we now present results for
the total cross sections $\sigma_{\pi N\rightarrow \pi\pi N}$ and
the invariant mass distributions $d\sigma/dM_{\pi N}$ or $d\sigma/dM_{\pi \pi}$
for all possible final $\pi\pi N$ states of $\pi^\pm p$ 
reactions 
in the energy region from threshold to invariant mass $W= 2$ GeV.
As mentioned in section I, our investigation thus is much more extensive
than all of the previous dynamical calculations of
$\pi N \rightarrow \pi\pi N$ reactions in both the
energy range covered and the $N^*$ states considered.

We can predict the
$\pi N \rightarrow \pi\pi N$ cross sections using
the information generated from the
 JLMS model except that due to the direct production
interaction $v_{\pi N,\pi\pi N}$ in Eq.(\ref{eq:tpipin-dir}).
This term, which can induce the 
$\pi N \rightarrow \pi\pi N \rightarrow \pi N$ mechanism to
influence the $\pi N$ elastic scattering, was not included
in the development of JLMS model. We thus first examine
the importance of this term. As discussed in section II, 
the contributions from
$v_{\pi N,\pi\pi N}$, calculated by using $T^{\text{dir}}_{\pi N, \pi\pi N}$
defined in Eq.(\ref{eq:dir-app-2}), is completely fixed by
 using the same parameters from JLMS model. Thus no 
additional
parameters are introduced  in our calculations.

 Our results for the 
total cross sections are shown 
in Fig.\ref{fig:jlms-pipp}. The
solid curves are from our full calculations and the dashed curves are from
turning off the term $T^{\text{dir}}_{\pi N, \pi\pi N}$.
We see that
both the magnitudes and the energy-dependence of the data for
all five two-pion production processes  can
be reproduced to a very large extent by our full calculations (solid curves).
Clearly, 
the direct $\pi N \rightarrow \pi\pi N$ mechanisms
play a significant role in determining the predicted
cross sections. In particular, it is instrumental in obtaining
 the agreement with the $\pi^+ p \rightarrow \pi^+\pi^+ n$
 data. Its effects  at low W can be  more clearly seen in Fig.~\ref{fig:jlms-pipp-th}. Here we also see
that the agreement with the data of $\pi^- p \rightarrow \pi^0\pi^-p$
is mainly due to the effects of $v_{\pi N,\pi\pi N}$.
 We  note here that
our full calculations (solid curves) in
Fig.\ref{fig:jlms-pipp-th} are comparable
to those of  the chiral perturbation theory calculation
of Refs.~\cite{bernard97,fettes00,mobed05}. 
This suggests that the model $v_{\pi N,\pi\pi N}$
considered here is fairly reasonable and
the discrepancies with the data
in the higher W region, seen in Fig.\ref{fig:jlms-pipp},
are more likely from the uncertainties in the
contributions from $\pi\Delta, \sigma N, \rho N$ transitions.

\begin{table}
\begin{ruledtabular}
\begin{tabular}{c|c|c|c|c|c|c}
\multicolumn{7}{c}{ }\\
\hline
      &       &\multicolumn{5}{c}{}\\
States&W (MeV)&\multicolumn{5}{c}{$\frac{\sigma_{MB}}{\sigma_{\text{total}}}(\%)$}\\
      &       &\multicolumn{5}{c}{}\\
\hline
&& $\pi N$ & $\eta N$ & $\pi\Delta$ & $\sigma N$ & $\rho N$    \\
\hline
$S_{11}$  & 1535  & 50.7 & 34.9 & 8.0 & 6.1 & 0.3  \\
&&&&&&\\
          & 1650 & 72.3 & 4.0 & 1.1 &  8.2 & 14.4  \\
&&&&&&\\
$S_{31}$  &1620   & 26.6 & 0 & 67.3 &0 &3.1   \\
\hline
\hline
$P_{11}$  & 1440  &62.5 &0 &7.7 &29.5 &0.3  \\
&&&&&& \\
          & 1710  &51.2 &2.0 &8.23 &24.6 &14.0 \\
&&&&&&\\

$P_{13}$  & 1720  &27.9 &0.2 &70.2 &0.9 &0.8  \\
&&&&&&\\
$P_{31}$  & 1910  &79.5 &0 &6.0 &0 &14.4  \\
&&&&&&\\ 
$P_{33}$  &1232 & 100 & 0 & 0 &0  &0  \\
&&&&&&\\
  &1600 & 76.4 & 0 & 13.6 &0  &10.0   \\
\hline
\hline
$D_{13}$ & 1520   &58.0 & 0.0 &36.8 &4.5 &0.8 \\
&&&&&&\\
$D_{15}$ & 1675&  38.6 &0.7 &56.5 &0.7 &3.5  \\
&&&&&&\\
$D_{33}$ &1700 & 12.4 &0 & 85.2&0 &2.4  \\
\hline
\hline
$F_{15}$ &1680 &   70.0 &0.0 &4.8 &19.3 &5.8  \\
&&&&&&\\
$F_{35}$ & 1905 & 10.9 &0 & 51.1& 0& 38.0 \\
&&&&&&\\
$F_{37}$ & 1950 &  39.8 &0 & 59.7&0 &0.5  \\
&&&&&&
\end{tabular}
\caption{Branching ratios of the             
$\pi N \rightarrow \pi N, \eta N, \pi \Delta, \sigma N, \rho N$ 
partial wave cross sections calculated from the resonant amplitude
Eq.(\ref{eq:tmbmb-r}).\label{tab:nstar}
$W$ is the total energy.}
\end{ruledtabular}
\end{table}

\begin{figure}[tb]
\centering
\includegraphics[clip,width=15cm]{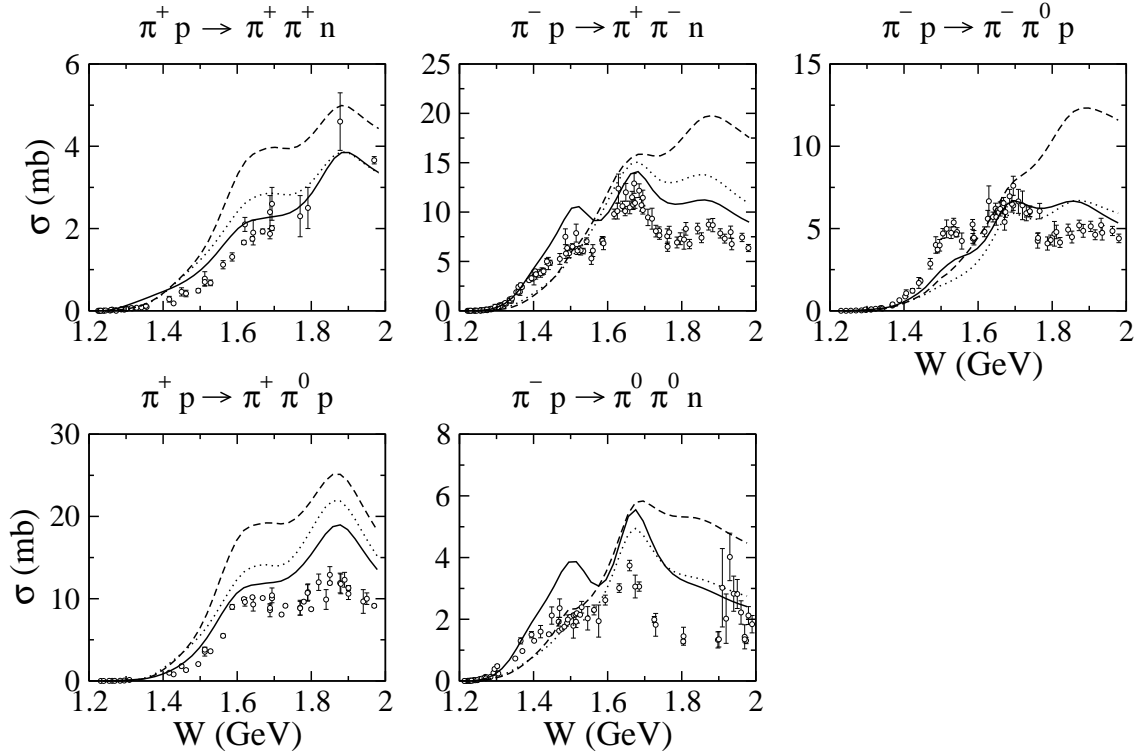}
\caption{The coupled-channels effects on $\pi N \rightarrow \pi\pi N$
reactions. The solid curves are from full calculations, the dotted curves 
are from keeping 
only $M'B'= MB$ in the Eq.(\ref{eq:no-cc}) and
in $\bar{\Gamma}_{N^* \rightarrow MB}$ of
Eqs.(\ref{eq:mb-nstar})-(\ref{eq:nstar-mb}),
the dashed curves are from setting $t_{MB,M'B'}=v_{MB,M'B'}$.
The data are from~\cite{arndt}. }
\label{fig:cc-tot}
\end{figure}

The main feature of this investigation is a dynamical
 coupled-channels treatment of $\pi\pi N$ channel which has the 
$\pi\Delta, \rho N, \sigma N$ resonant channels.
In our calculation, this effect can be explicitly seen by writing 
the coupled-channels equations, Eq.(\ref{eq:cc-mbmb}), as
\begin{eqnarray}
t_{\pi N,MB}(E) = \sum_{M^{'}B^{'}}[1-vG]^{-1}_{\pi N,M^{'}B^{'}}
v_{M^{'}B^{'}, MB}
\label{eq:no-cc}
\end{eqnarray}
where $MB=\pi\Delta, \rho N, \sigma N$, and
the intermediate meson-baryon states can be 
$M^{'}B^{'}=\pi N,\eta N, \pi\Delta, \sigma N, \rho N$. When only
the term with $M'B'= MB$ in the Eq.(\ref{eq:no-cc}) and
in $\bar{\Gamma}_{N^* \rightarrow MB}$ of 
Eqs.(\ref{eq:mb-nstar})-(\ref{eq:nstar-mb}) is kept, the
calculated total cross sections (full curves) are changed to the
dotted curves in Fig.\ref{fig:cc-tot}.
 If we further neglect the coupled-channels
effects 
by setting $t_{\pi N, MB}=v_{\pi N, MB}$, we then get the
dashed curves which are very different from the full calculations (solid curves),
in particular in the high $W$ region. 

\begin{figure}[tb]
\centering
\includegraphics[clip,width=12cm]{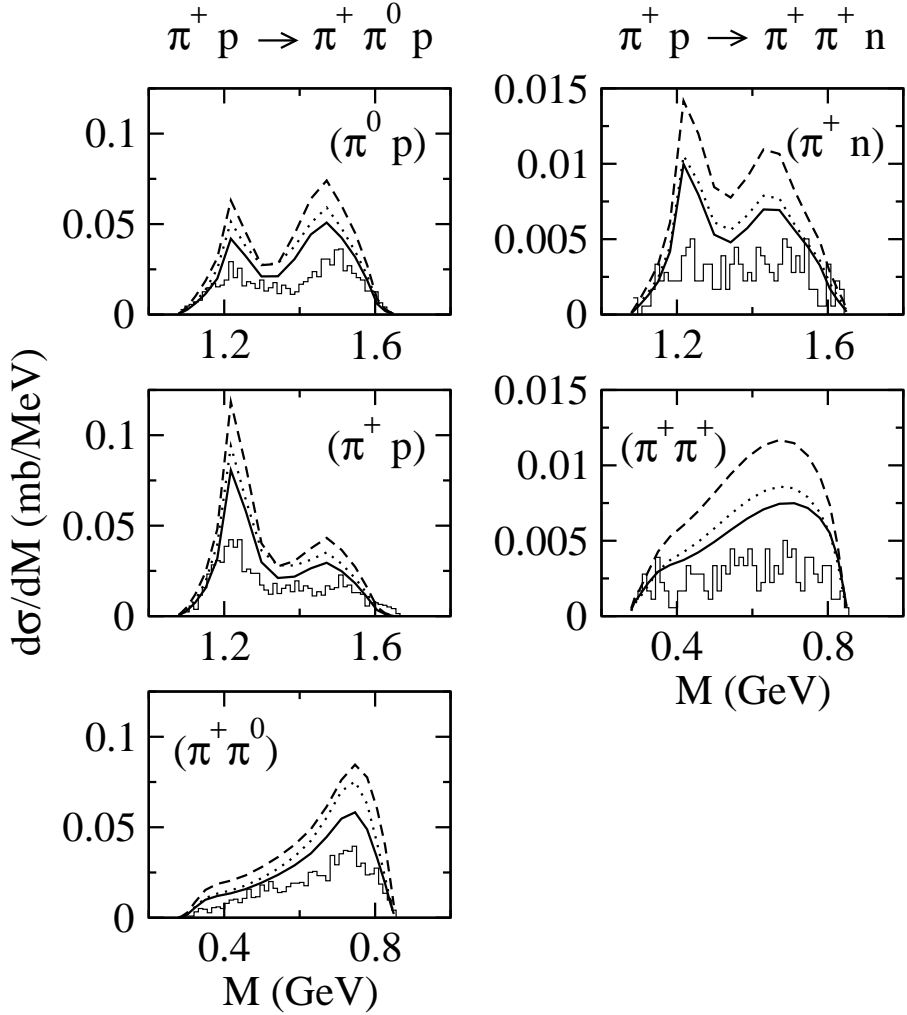}
\caption{Coupled-channels effects on the invariant mass distributions
of $\pi^+ p \rightarrow \pi\pi N$ at $W = 1.79 $ GeV. 
The solid curves are from full calculations, the dotted curves
are from keeping
only $M'B'= MB$ in the Eq.(\ref{eq:no-cc}) and
in $\bar{\Gamma}_{N^* \rightarrow MB}$ of
Eqs.(\ref{eq:mb-nstar})-(\ref{eq:nstar-mb}),
the dashed curves are from setting $t_{MB,M'B'}=v_{MB,M'B'}$. 
The data are from R. Arndt~\cite{arndt}. } \label{fig:cc-pipp}
\end{figure}

\begin{figure}[t]
\centering
\includegraphics[clip,width=12cm]{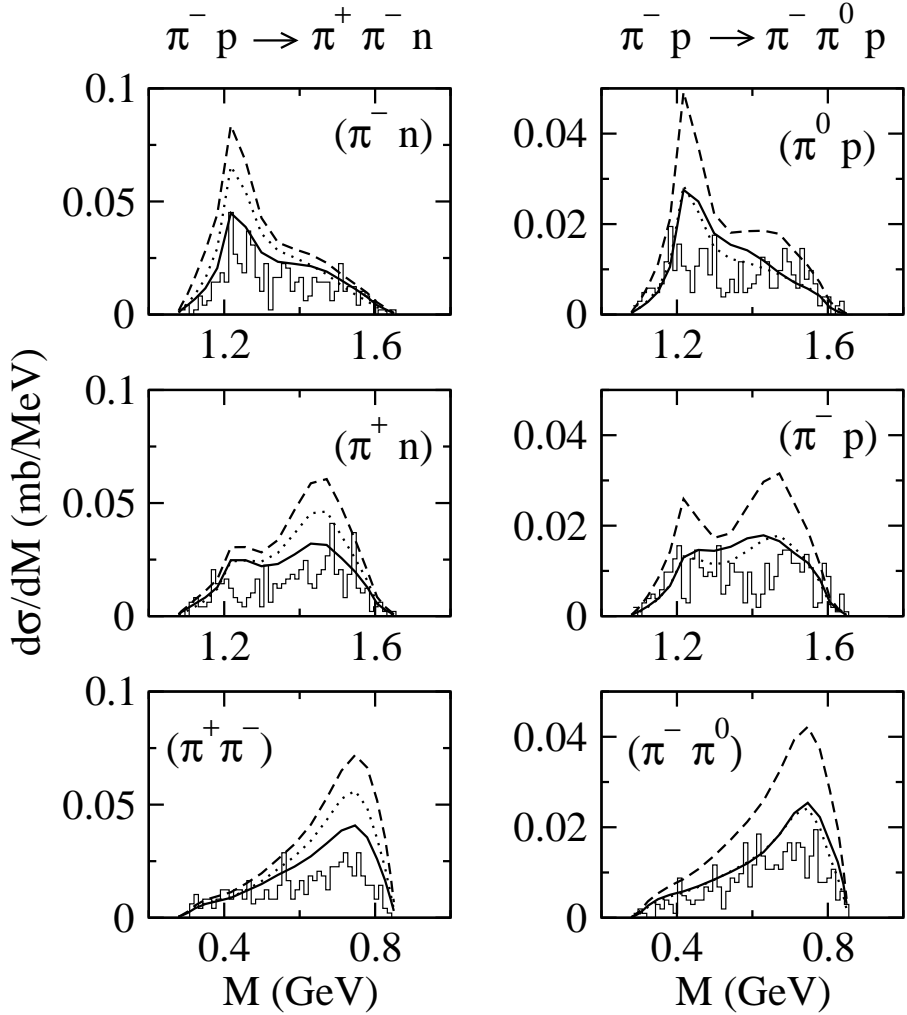}
\caption{Coupled-channels effects on the invariant mass distributions
of $\pi^- p \rightarrow \pi\pi N$ at $W = 1.79 $ GeV.
The solid curves are from full calculations, the dotted curves
are from keeping
only $M'B'= MB$ in the Eq.(\ref{eq:no-cc}) and
in $\bar{\Gamma}_{N^* \rightarrow MB}$ of
Eqs.(\ref{eq:mb-nstar})-(\ref{eq:nstar-mb}),
the dashed curves are from setting $t_{MB,M'B'}=v_{MB,M'B'}$. The 
data are from R. Arndt~\cite{arndt}. } \label{fig:cc-pimp}
\end{figure}

To see the coupled-channels effects more clearly, we show the corresponding
results for  the $\pi N$ and $\pi\pi$ invariant mass
distributions at $W=1.79 $ GeV in 
Figs.\ref{fig:cc-pipp}-\ref{fig:cc-pimp}.
Our full calculations (solid curves) are able to reproduce the main features
of the data. Comparing them with the dotted and dashed curves, it is
clear that the coupled-channels effects can change strongly both
the magnitudes and energy-dependence of the $\pi N \rightarrow \pi\pi N$
cross sections. 

\begin{figure}[tbh]
\centering
\includegraphics[clip,width=12cm]{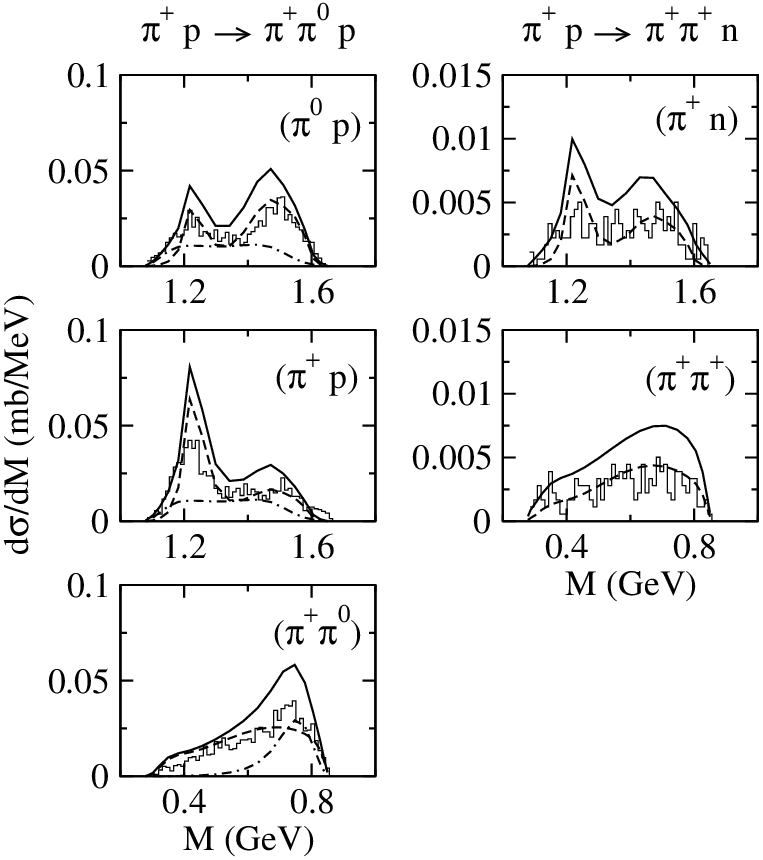}
\caption{Contributions from $\pi \Delta$ (dashed) and
$\rho N$ (dot-dashed) channels to the invariant mass distributions of
 $\pi^+ p \rightarrow \pi\pi N$ at $W = 1.79 $ GeV. The data are from 
R. Arndt~\cite{arndt}.}
\label{fig:ch-pipp}
\end{figure}

\begin{figure}[tbh]
\centering
\includegraphics[clip,width=12cm]{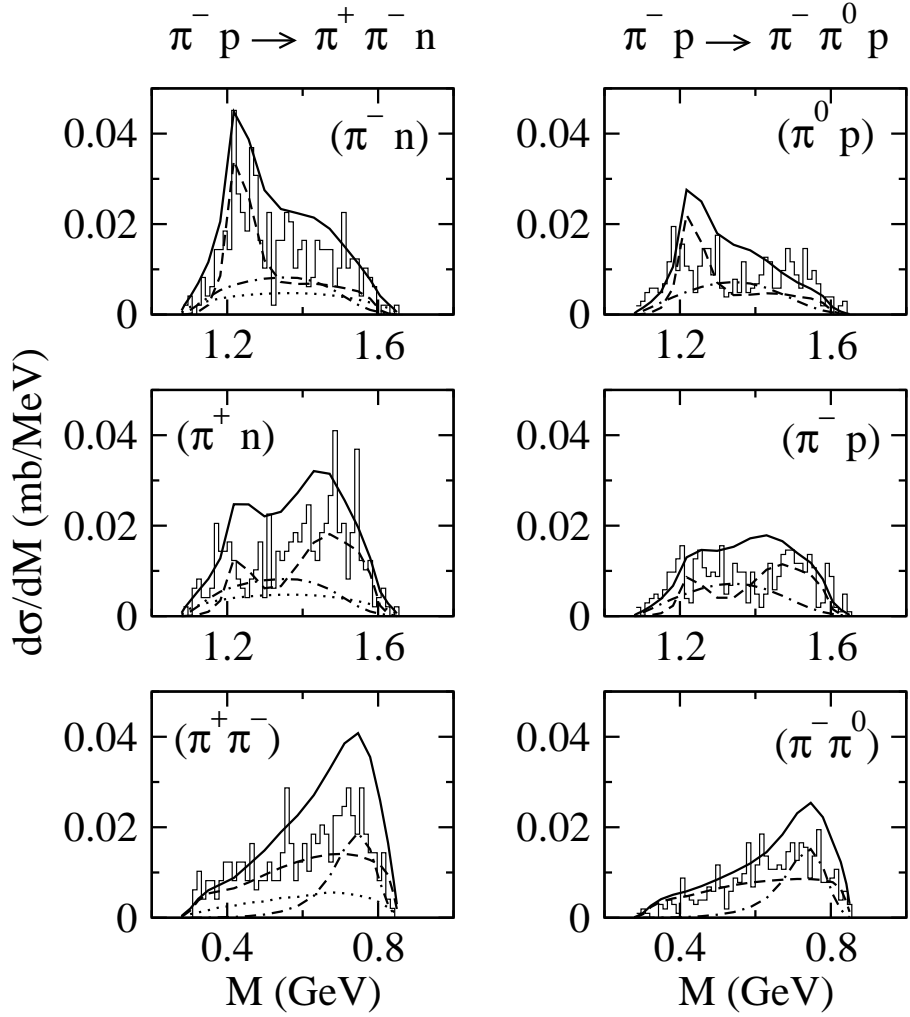}
\caption{Contributions from $\pi \Delta$ (dashed), $\sigma N$ (dotted) and
$\rho N$ (dot-dashed) channels to the invariant mass distributions of
 $\pi^- p \rightarrow \pi\pi N$ at $W = 1.79 $ GeV. The data are from 
R. Arndt~\cite{arndt}.}
\label{fig:ch-pimp}
\end{figure}

To further see the dynamical content of our model, we show in 
Figs.\ref{fig:ch-pipp}-\ref{fig:ch-pimp} the contributions to the
invariant mass distributions at $W=1.79 $ GeV from
each of the processes via the final 
$MB=\pi\Delta$, $\rho N$, $\sigma N$ defined by 
Eqs.(\ref{eq:tpipin-pid})-(\ref{eq:tpipin-sigman}).
The results shown in Figs.\ref{fig:ch-pipp}-\ref{fig:ch-pimp}
indicate that the full coupled-channels calculations (solid curves) involve
rather complex interference effects between these three unstable particle 
channels.
To improve the model, we need to tune their relative importance.

\begin{figure}[h]
\centering
\includegraphics[clip,width=12cm]{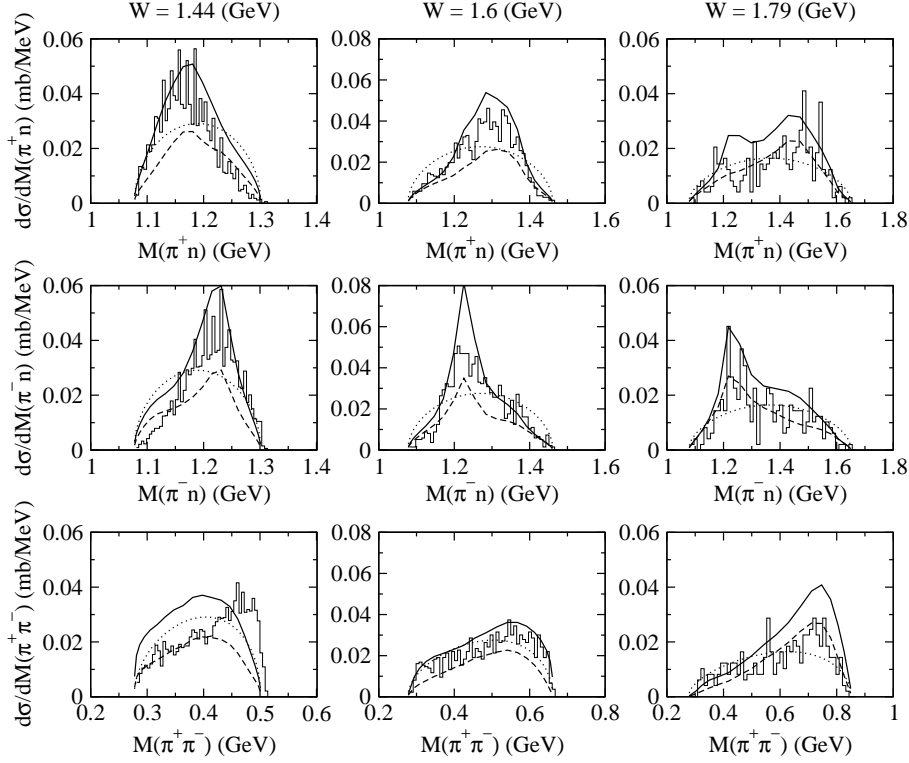}
\caption{The invariant mass distributions of
$\pi^- p \rightarrow \pi^+\pi^- n$ at $W = 1.44, 1.60, 1.79$ GeV.
The solid curves are the full calculations, the dashed curves are from 
non-resonant amplitudes, and the dotted curves are the phase-space
normalized to the data. The data are from Ref.~\cite{arndt}. }
\label{fig:pm-3w}
\end{figure}

Compared with all previous investigations, another feature of this 
investigation is our treatment of the $N^*$ resonance amplitudes.
These amplitudes are generated from 16 bare states, as given
in Ref.~\cite{jlms07}, which are dressed
by the non-resonant interactions, as required by the unitarity condition
and defined by Eq.(\ref{eq:tmbmb-r})-(\ref{eq:nstar-mb}).
In Fig.~\ref{fig:pm-3w}, we compare the full results (solid curves) and 
that calculated from keeping only the non-resonant
amplitudes (dashed curves) for the invariant mass distributions
of $\pi^- p \rightarrow \pi^+\pi^- n$ at $W= 1.44$, $1.60$, $1.79$ GeV.
Here we note that 
the peaks of dashed curves in $M(\pi^+ n)$ and $M(\pi^- n)$ distributions
around 1.2 GeV in Fig.~\ref{fig:pm-3w} are due to the decay of $\Delta$
in the intermediate $\pi\Delta$ state of the ``non-resonant processes''
of $\pi N\to \pi\Delta\to\pi\pi N$, whose amplitude is defined by
Eq.~(\ref{eq:tpipin-pid}) but replacing $T_{\pi N,\pi\Delta}$
with its non-resonant amplitude $t_{\pi N,\pi\Delta}$ generated from
Eq.(\ref{eq:cc-mbmb}).
Similarly, the decay of $\rho$ ($\sigma$) in the intermediate $\rho N$
($\sigma N$) of the ``non-resonant processes'' of
$\pi N\to \rho N(\sigma N)\to\pi\pi N$ can be responsible
for the peaks of dashed curves in the $M(\pi^+\pi^-)$ distributions
(lower row of Fig.~\ref{fig:pm-3w}).

By comparing the solid and dashed curves in Fig.~\ref{fig:pm-3w}, 
it is clear that
the full calculations involve comparable contributions from
resonant and non-resonant amplitudes.
In the same figure, we also show $\pi \pi N$ phase-space 
distributions (dotted curves) normalized to data. 
The shapes of both theoretical results deviate
significantly from the phase-space.

The results shown in Fig.~\ref{fig:pm-3w}
indicate that the fits to the data depend strongly on the parameters associated
with the $N^* \rightarrow \pi\Delta$, $\rho N$, $\sigma N$ vertex functions
which are treated purely phenomenologically within JLMS model. 
One possible improvement
of the model is to explore how these vertex functions can be calculated
from sound hadron structure calculations. An attempt along this line for
a two-channel $\pi N$ scattering in $S_{11}$ state was
pursued in Ref.~\cite{yoshimoto} using the constituent quark model, but was not
successful.

The complexity of the calculated resonant amplitudes
can be further seen  in Table I
where we show the calculated
branching ratios of the contributions from each
channel to the partial-wave cross sections calculated from
the resonant amplitude $t^R_{\pi N, MB}$
(Eq.(\ref{eq:tmbmb-r})) at the resonant energies listed by Particle Data Group.
Clearly, the resonant amplitudes  
involve strong interference between the
$ \pi N\rightarrow \pi\Delta, \rho N, \sigma N \rightarrow \pi\pi N$ amplitudes.
To improve the model, we need to tune their relative importance.
Clearly more detailed
data, such as the single or double angular distributions and polarizations,
are needed to make significant progress. 
We emphasize here that the results listed in Table I
 are {\it not} the branching ratios of
$N^*\rightarrow MB$ decay widths at
the resonance poles which will be extracted from using the analytic
continuation methods developed in Ref.~\cite{ssl08}.
These results just give some ideas about the relative importance between
different channels at some energies listed by PDG.

\begin{figure}[hbt]
\centering
\includegraphics[clip,width=6cm]{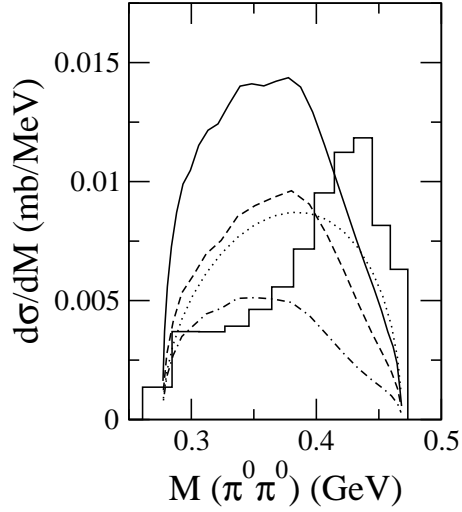}
\caption{The invariant mass distributions of
$\pi^- p \rightarrow \pi^0\pi^0 n$ at $W=1.40$ GeV. 
The solid curve is the full calculation.
The dashed (dot-dashed) curve is from keeping only the non-resonant
(resonant) amplitude in the calculations.
The dotted curve is the phase-space normalized to data.
The data are from Ref.~\cite{crystal}.
(The data is transformed to $d\sigma/dM$ using the relation
$d\sigma/dM = 2M d\sigma/dM^2$.)}
\label{fig:pi0pi0n-1400}
\end{figure}

In Fig.~\ref{fig:pm-3w}, we also observe that our predictions
do poorly in describing the $\pi^+\pi^-$ distribution at low 
$W= 1.44 $ GeV. We have found that this is the case for all two-pion
invariant mass distributions of $\pi^-p\to\pi^+\pi^-n$ and 
$\pi^-p\to\pi^0\pi^0n$ reactions at low $W \lesssim$ 1.5 GeV.
This is given in more detail in Fig.~\ref{fig:pi0pi0n-1400} for
$\pi^- p \rightarrow \pi^0\pi^0 n$ reaction.
We see that 
our prediction (solid curve) does not reproduce the
data from  the Crystal Ball collaboration~\cite{crystal}. 
We also show the results
from keeping only the non-resonance amplitudes (dashed curve) and
only the resonant amplitude (dot-dashed curve).
The shapes of all theoretical curves
are similar to phase-space (dotted curve) and are far from the
data. 
We have found that the problem can not be easily resolved by simply adjusting
$N^*$ parameters, in particular those in the most controversial $P_{11}$
partial waves. 
It requires detailed analysis and  more extensive $\pi N \rightarrow \pi\pi N$
data to resolve the problem.

\begin{figure}[hbt]
\centering
\includegraphics[clip,width=15cm]{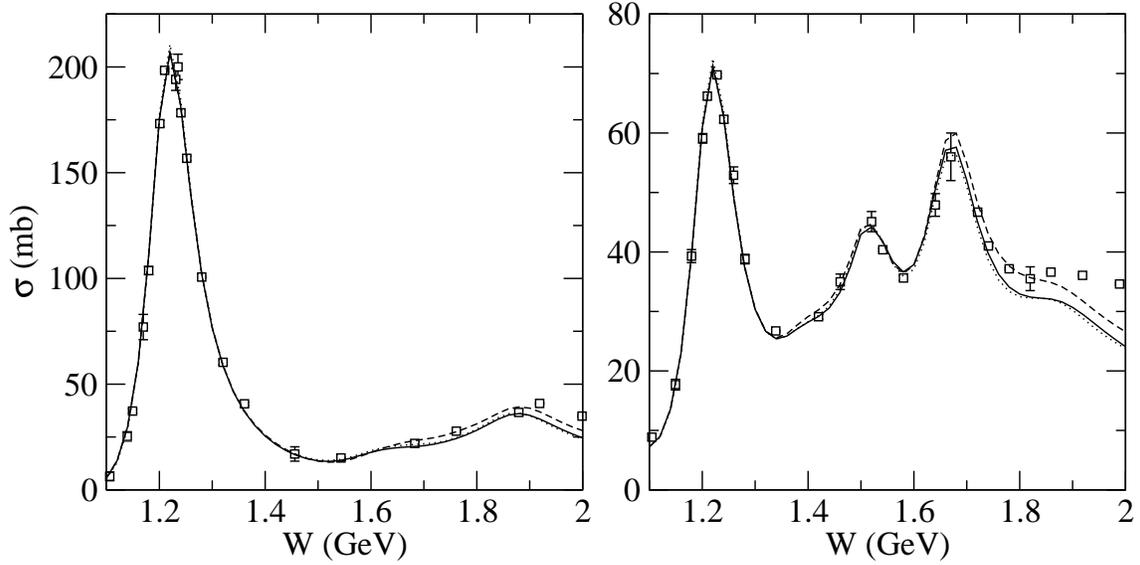}
\caption{Total cross sections of $\pi^+ p$ (left) and $\pi^- p$ reactions.
Solid curves: using the optical theorem
$\sigma^{(\text{tot})}_{\text{opt}} = (2\pi/k)F_{\pi N,\pi N}(\theta=0)$.
Dotted curves:
$\sigma^{(\text{tot})} = \sigma_{\pi N, \pi N} + \sigma_{\pi N, \pi \eta}
+ \sigma_{\pi N, \pi\pi N}$ (no $T^{\text{dir}}_{\pi N,\pi\pi N}$).
Dashed curves: $\sigma^{(\text{tot})} = \sigma_{\pi N, \pi N}
 + \sigma_{\pi N, \pi \eta}
+\sigma_{\pi N, \pi\pi N}$. Solid and dotted curves are not distinguishable. 
Only few data are shown for a clear comparison between curves.
The data are from Refs.~\cite{pdg,cns}.
}
\label{fig:v23effect} 
\end{figure}

\section{Summary}
In this paper we have shown that the predictions from the JLMS model can
describe to a rather large extent 
the data of  total cross sections and $\pi N$ and $\pi\pi$ invariant
mass distributions  of $\pi N \rightarrow \pi\pi N$ reactions in the
energy region from threshold to $W= 2$ GeV. Our investigation is thus more extensive than all 
previous dynamical calculations of this reaction in both the energy range
covered and the $N^*$ states considered.

We have demonstrated the importance of the coupled-channels effects and
strong interference between the $\pi\Delta$, $\rho N$, and $\sigma N$.
The problem in identifying the mechanisms for
improving the considered JLMS model is further complicated
by the finding that the contributions from resonant and non-resonant
amplitudes are comparable.

An important finding in this work is that the direct $v_{\pi N,\pi\pi N}$
mechanisms, illustrated in Fig.~\ref{fig:mbpipin}, play a significant role
in obtaining good agreement with the data, especially in the $W \leq $ 1.4 GeV
region where our results are comparable to those from the
chiral perturbation theory calculations~\cite{bernard97,fettes00,mobed05}.
This raises the question on the extent to which our elastic scattering results
will be changed by the effect due to $v_{\pi N,\pi\pi N}$ and how
the unitarity condition is violated. For the former one, it 
can not be answered easily since it will involve solving
three-body
$\pi\pi N \rightarrow \pi\pi N$ scattering equations, 
as discussed in Ref.~\cite{msl07}.
We however can examine the unitarity condition by comparing
the total cross sections calculated from
(a)  using the optical theorem
$\sigma^{(\text{tot})}_{\text{opt}} = (2\pi/k)F_{\pi N,\pi N}(\theta=0)$, (b)
$\sigma^{(\text{tot})} = \sigma_{\pi N, \pi N} + \sigma_{\pi N, \pi \eta}
+ \sigma_{\pi N, \pi\pi N}$ (no $T^{\text{dir}}_{\pi N,\pi\pi N}$),
(c) $\sigma^{(\text{tot})} = \sigma_{\pi N, \pi N} + \sigma_{\pi N, \pi \eta}
+\sigma_{\pi N, \pi\pi N}$. These are shown in Fig.\ref{fig:v23effect}.
Here, $\sigma_{\pi N, \pi N}$ and $\sigma_{\pi N, \pi \eta}$ 
are directly calculated from the employed amplitudes generated from JLMS model
within which the effects of $v_{\pi N,\pi\pi N}$ are not included, 
$\sigma_{\pi N, \pi\pi N}$ (no $T^{\text{dir}}_{\pi N,\pi\pi N})$ and
$\sigma_{\pi N, \pi\pi N}$ are calculated from using the formula given in 
section II and III, with $T^{\text{dir}}_{\pi N,\pi\pi N}$ defined by
$v_{\pi N,\pi\pi N}$ through Eq.(16). 
We see that (a) (solid curves) and (b) (dotted curves)
agree completely as required by the unitarity condition within the JLMS model.
 Their
differences with  (c) (dashed curves) measure the violation of
the unitarity condition when the effects due to $v_{\pi N,\pi\pi N}$ are not
consistently included in solving the coupled-channels scattering equations.
Clearly, the unitarity condition is 
violated significantly mainly in the high W region. For example, 
the results for $\pi^- p$ total cross sections at $W=1.8$ GeV
are: (a)=32.94 mb, (b)=32.48mb, and (c)=35.57 mb.
However, the results shown in
Fig.\ref{fig:v23effect} just mean that the effect of $v_{\pi N,\pi\pi N}$ 
will not
change significantly the elastic differential cross sections at $forward$
 angles.
For a complete unitary calculations for all $\pi N$ reaction observables,
 we need to
include $v_{\pi N,\pi\pi N}$ effects in solving the coupled-channels
equations, as detailed in Ref.\cite{msl07}.
This is being pursued along with our effort in developing a combined fit
to the world data of $\pi N, \gamma N \rightarrow \pi N, \eta N, \pi\pi N$.
Our progress in this direction will be reported elsewhere.

Our analysis presented  in Figs.~\ref{fig:cc-pipp}-\ref{fig:pi0pi0n-1400}
indicates
the complication of the $\pi N \rightarrow \pi\pi N$ problem. 
To improve our model, more experimental data, such as the single 
or double angular distributions and polarization observables, are 
needed to pin down the parameters of the model.
With the recent effort~\cite{arndt}, progress in this direction could be
realized in the near future. Of course,  experimental efforts at the
new hadron facilities such as JPARC are highly desirable.

\clearpage

\begin{acknowledgments}
We would like to thank R. Arndt for recovering the old data of
$\pi N \rightarrow \pi\pi N$ reactions.
This work is supported by the U.S. Department of Energy, 
Office of Nuclear Physics Division, under contract No. DE-AC02-06CH11357, 
and Contract No. DE-AC05-060R23177 
under which Jefferson Science Associates operates Jefferson Lab,
and by the Japan Society for the Promotion of Science,
Grant-in-Aid for Scientific Research(c) 20540270. This work is also
partially supported by Grant No. FIS2005-03142 from MEC (Spain) 
and FEDER and European Hadron Physics Project RII3-CT-2004-506078. 
The computations were performed at NERSC (LBNL) and Barcelona 
Supercomputing Center (BSC/CNS) (Spain). The authors thankfully acknowledges 
the computer resources, technical expertise and assistance provided by the 
Barcelona Supercomputing Center - Centro Nacional de Supercomputacion (Spain).
\end{acknowledgments}

\clearpage

\appendix
\section{Self Energy in unstable $MB$ propagators}
\label{app1}

In this appendix we give a derivation of Eq.(35).
To be more explicit, let us consider
$eR = \pi\Delta$ for Eq.(35) and suppress spin-isospin indices. 
The starting point is Eq.(21) of 
Ref.\cite{msl07} which defines 
the formulation used in JLMS model and this work.

\begin{figure}[b]
\centering
\includegraphics[clip,width=6cm]{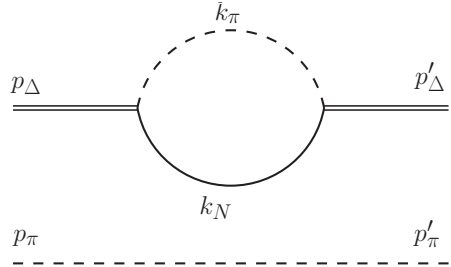}
\caption{ Graphical illustration of Eq.(A1) for calculating
the $\Delta$ self energy in $\pi\Delta$ propagator.
}
\label{fig:sig}
\end{figure}

Since the vertex interaction $H_I=\Gamma_{\Delta,\pi N}$ conserves
the total three momentum of
the system, we have
\begin{eqnarray}
& &\left[\langle \vec{p^\prime}_\Delta \vec{p^\prime}_\pi|H_I\frac{P_{\pi\pi N}}
{E-K_\pi-K_\pi-K_N+i\epsilon}H_I 
|\vec{p}_\Delta \vec{p}_\pi\rangle\right]_{\text{un-connected}} 
\nonumber \\
&& =\delta(\vec{p^\prime}_\Delta -\vec{p}_\Delta)
\delta(\vec{p^\prime}_\pi-\vec{p}_\pi)\Sigma_{\pi\Delta}(p_\pi,E).
\end{eqnarray}
The kinematics for evaluating Eq.(A1) is illustrated in Fig.\ref{fig:sig}.
To proceed further, we then use the following well known relativistic
kinematic relations (for example, see Ref.\cite{tabakin} and
section 2.3 of Ref.\cite{keisder})
\begin{eqnarray}
\vec{P} &=& \vec{k}_\pi + \vec{k}_N, \\
\vec{q}&=&\frac{1}{2M_{\pi N}(E_{\pi N}+M_{\pi N})}\nonumber \\
& &\times [(M^2_{\pi N}+2E_N(\vec{k}_N)M_{\pi N}+
m^2_N-m^2_\pi)\vec{k}_\pi
-(M^2_{\pi N}+2E_{\pi}(\vec{k}_\pi)M_{\pi N}+m^2_\pi-m^2_N)\vec{k}_N],
\nonumber \\
\end{eqnarray}
with
\begin{eqnarray}
E_{\pi N}&=&E_\pi(\vec{k}_\pi)+E_N(\vec{k}_N), \nonumber \\
M_{\pi N}&=&E_\pi(\vec{q})+E_N(\vec{q}) \nonumber \\
&=&[E_{\pi N}^2-\vec{P}^{\,\,2}]^{1/2}. 
\end{eqnarray}
We then have
\begin{eqnarray}
d\vec{k}_\pi d\vec{k}_N = \left|\frac{\partial(\vec{k}_\pi, \vec{k}_N)}
{\partial(\vec{P}, \vec{q})}\right|d\vec{P} d\vec{q},
\end{eqnarray}
with
\begin{eqnarray}
\frac{\partial(\vec{k}_\pi, \vec{k}_N)}
{\partial(\vec{P}, \vec{q})} 
&=& \frac{E_\pi(\vec{k}_\pi)E_N(\vec{k}_N)}{E_\pi(\vec{k}_\pi)+E_N(\vec{k}_N)} 
\cdot \frac{E_\pi(\vec{q})+E_N(\vec{q})}
{E_\pi(\vec{q})E_N(\vec{q})} \nonumber \\
&=& \frac{E_\pi(\vec{k}_\pi)E_N(\vec{k}_N)}{E_\pi(\vec{q})E_N(\vec{q})}
\cdot \frac{ M_{\pi N}(q)}
{[M^2_{\pi N}(q) + \vec{P}^{\,\,2}]^{1/2}}.
\end{eqnarray}
With the Lorentz invariance property Eq.(27), we can calculate 
the matrix element of $H_I$ in terms of that in the 
rest frame of $\Delta$
\begin{eqnarray}
\langle\vec{k}_\pi \vec{k}_N |H_I|\vec{p}_\Delta\rangle
=\delta(\vec{p}_\Delta -\vec{k}_\pi- \vec{k}_N)
\sqrt{\frac{E_\pi(q)E_N(q)m_\Delta}{E_\pi(k_\pi)E_N(k_N)E_\Delta(p_\Delta)}}
\langle\vec{q}, -\vec{q} |H_I|\vec{0}\rangle. 
\end{eqnarray}
By using the above relations, we then have
\begin{eqnarray}
& &\langle \vec{p^\prime}_\Delta \vec{p^\prime}_\pi|H_I\frac{P_{\pi\pi N}}
{E-K_\pi-K_\pi-K_N+i\epsilon}H_I |\vec{p}_\Delta \vec{p}_\pi\rangle 
\nonumber \\
&& =\delta(\vec{p^\prime}_\pi-\vec{p}_\pi)
\int  \langle\vec{p^\prime}_\Delta |H_I|\vec{k}_\pi \vec{k}_N \rangle
 \frac{d\vec{k}_\pi d\vec{k}_N}{E-E_\pi(\vec{p}_\pi)
-E_\pi(\vec{k}_\pi) - E_N(\vec{k}_N) + i\epsilon} 
 \langle\vec{k}_\pi \vec{k}_N |H_I|\vec{p}_\Delta\rangle. \nonumber \\
\end{eqnarray}
By using Eqs.(A5)-(A7), we then obtain
\begin{eqnarray}
& &\langle \vec{p^\prime}_\Delta \vec{p^\prime}_\pi|H_I\frac{P_{\pi\pi N}}
{E-K_\pi-K_\pi-K_N+i\epsilon}H_I |\vec{p}_\Delta \vec{p}_\pi\rangle \nonumber \\
&& = \delta(\vec{p^\prime}_\pi-\vec{p}_\pi)
\int
[\frac{E_\pi(q)E_N(q)m_\Delta}{E_\pi(k_\pi)E_N(k_N)E_\Delta(p_\Delta)}]
|\langle\vec{q}, -\vec{q} |H_I|\vec{0}\rangle|^2
\delta(\vec{p^\prime}_\Delta-\vec{P})\delta(\vec{p}_\Delta-\vec{P})
\nonumber \\
& & \times [\frac{E_\pi(\vec{k}_\pi)E_N(\vec{k}_N)}
{E_\pi(\vec{q})E_N(\vec{q})}\cdot \frac{ M_{\pi N}(q)}
{[M^2_{\pi N}(q) + \vec{P}^{\,\,2}]^{1/2}}]
\frac{d\vec{q} d\vec{P}}
{E-E_\pi(\vec{p}_\pi)
-\{[E_\pi(\vec{q})+E_N(\vec{q})]^2+ \vec{P}^{\,\,2}\}^{1/2} + i\epsilon} \nonumber \\
&& =\delta(\vec{p^\prime}_\pi-\vec{p}_\pi) \delta(\vec{p^\prime}_\Delta
-\vec{p}_\Delta)\frac{m_\Delta}{E_\Delta(p_\Delta)} \nonumber \\
& &\times \int \frac{ M_{\pi N}(q)}{[M^2_{\pi N}(q) + \vec{P}^{\,\,2}]^{1/2}}
\frac{d\vec{q}|\langle\vec{q}, -\vec{q} |H_I|\vec{0}\rangle|^2}
{E-E_\pi(\vec{p}_\pi)
-\{[E_\pi(\vec{q})+E_N(\vec{q})]^2 + \vec{p}_\Delta^{\,\,2}\}^{1/2} + i\epsilon}. 
\end{eqnarray}
Comparing Eqs.(A9) and (A1) and using the partial wave expansion
Eq.(28), we then obtain in the center of mass frame 
($\vec{p}_\pi = -\vec{p}_\Delta= \vec{k}$)
\begin{eqnarray}
\Sigma_{\pi\Delta}(k,E) =\frac{m_\Delta}{E_\Delta(k)}
\int q^2 dq \frac{ M_{\pi N}(q)}{[M^2_{\pi N}(q) + k^{\,\,2}]^{1/2}}
\frac{|f_{\Delta \rightarrow \pi N}|^2}
{E-E_\pi(\vec{k})
-\{[E_\pi(\vec{q})+E_N(\vec{q})]^2 + {k}^{\,\,2}\}^{1/2} + i\epsilon}.
\nonumber\\
\end{eqnarray}
Eq.(A10) is Eq.(35) for $eR=\pi\Delta$.

\end{document}